\documentclass[a4paper,12pt]{article}
\usepackage[utf8x]{inputenc}
\usepackage{amsfonts}
\usepackage{amsmath}
\usepackage{amssymb}
\usepackage{amsthm}
\usepackage{booktabs}
\usepackage{multirow}
\usepackage{graphicx} 
\usepackage{graphics}
\usepackage[round]{natbib}
\usepackage{hyperref}
\usepackage{longtable}
\usepackage[]{threeparttable}
\usepackage{rotating}
\usepackage{xcolor}
\usepackage{comment}

  \newcommand{\clovis}{\color{red}}

\def\tr{{\rm T}}

\def\diag{{\rm diag}}
\def\tr{{\rm tr}}

\newcommand{\R}{{\rm I}\negthinspace {\rm R}}

\newcounter{example}

\title{Efficient implementation of median bias reduction with applications to general regression models}
\author{E. C. KENNE PAGUI, A. SALVAN and N. SARTORI \\
\small Department of Statistical Sciences, University of Padova \\
\small eulogeclovis.kennepagui@unipd.it, alessandra.salvan@unipd.it,  nicola.sartori@unipd.it
}
\date{}
\begin{document}

\maketitle

\begin{abstract}
\noindent
In numerous regular statistical models,  median bias reduction \citep{kenne2017} has proven to be a noteworthy improvement over maximum likelihood, alternative to mean bias reduction. The estimator is obtained as solution to a modified score equation ensuring  smaller asymptotic median bias than the maximum likelihood estimator.   This paper provides a simplified algebraic form of the adjustment term for general regular models. With the new formula, the estimation procedure benefits from a considerable computational gain by avoiding multiple summations and thus allows an efficient implementation.  More importantly, the new formulation allows to highlight how the median bias reduction adjustment  can be obtained by adding an extra term to the mean bias reduction adjustment.
Illustrations are provided through new applications of median bias reduction to two regression models not belonging to the generalized linear models class, extended beta regression and beta-binomial regression. Mean bias reduction is also provided here  for the latter model.  Simulation studies  show remarkable componentwise median centering of the median bias reduced estimator, while variability and coverage of related confidence intervals are comparable with those of mean bias reduction. Moreover, empirical results  for the beta-binomial model show that  the method is successful in solving maximum likelihood boundary estimate problem. 

\noindent

\end{abstract}

\noindent
\emph{Some key words:} Beta-binomial regression; Beta regression; Bias reduction; Boundary estimate; Maximum likelihood;   Median unbiasedness; Modified score.

%%%%%%%%%%%%%%%%%%%%%%%%%%%%%%
\section{Introduction}
%%%%%%%%%%%%%%%%%%%%%%%%%%%%%%
Consider estimation of a $p$-dimensional parameter $\theta$ in a regular parametric model based on a sample of size $n$. With moderate information, the maximum likelihood estimator can be highly inaccurate.
Several proposals have been developed to correct the estimate or the estimating function, with the latter approach having the advantage of not requiring the finiteness of the maximum likelihood estimate. 
Adjusted score functions for mean bias reduction were proposed by \citet{firth1993} and in  subsequent papers by \citet{kosmi2009, kosmi2010}. Median bias reduction was developed in  \citet{kenne2017},  and is such  that each component of the estimator has, to third-order, the same probability of underestimating and overestimating the corresponding parameter component.
Both mean and median bias reduction  consist  of adding a suitable adjustment term to the score function
and then solving the resulting adjusted score equation. 

While mean bias reduction is tied to a specific parameterization and only equivariance under linear transformations of the parameter is guaranteed, median bias reduction delivers estimators that are exactly invariant in terms of their improved median bias properties under  monotone component-wise transformations of the parameters.

    The median modified  score function  proposed by  \citet{kenne2017}  is obtained by adding an adjustment term of order $O(1)$ to the score function. Such adjustment  term was expressed in \citet{kenne2017} using index notation, which is a tool that allows to write complex formulae in a compact form, but is not necessarily optimal for their implementation.
%The estimator is defined as solution of the
%resulting estimating equation is median unbiased with a higher order of accuracy than the maximum 
%ikelihood estimator.
 In this paper, we give a new matrix expression of the adjustment term, similar to that of \citet[Section 3]{kosmi2010} for mean bias reduction.  The benefit is twofold: the new expression   allows a   more efficient implementation for general parametric models, and it also highlights a general connection  between mean and median bias reduction, thus  extending results  for generalized linear models in \citet{kosmidis2018mean}.

%The paper is structured as follows. Section 2 gives the new expression for the adjustment term. As an example, we show that the adjustment for generalized linear models in \citet{kosmidis2018mean} is obtained as a special case. 
Illustrations are provided for two regression models not belonging to the generalized linear models class. In particular, in Section 3  the new formulation is applied to  double index beta regression, while Section 4 is devoted to beta-binomial regression, where the expression for mean bias reduction is also derived.  For both models, R packages are available. All methods are  assessed and compared through  simulation experiments. The results confirm that the median bias reduced estimator succeeds in achieving componentwise median centering. Moreover,   both mean and median bias reduced are empirically found to solve the boundary estimate problem that may arise for maximum likelihood in beta-binomial regression.

%%%%%%%%%%%%%%%%%%%%%%%%%%%%%%
\section{Median modified score}
%%%%%%%%%%%%%%%%%%%%%%%%%%%%%%

Let us denote the vector parameter by $\theta=(\theta_1,\ldots,\theta_{p})^\top$ and  a generic component  of $\theta$ by  $\theta_r$. Let $ U_r=\partial \ell(\theta)/\partial\theta_r$ ($r=1,\ldots,p$) be the elements of the score vector  $U(\theta)$, where  $\ell(\theta)$ is the log likelihood function for $\theta$.
Let  $j(\theta) =-\partial U(\theta)/\partial\theta^\top$ be the observed  information matrix and  $i(\theta)=E\left\{j(\theta) \right\}$ be the Fisher information matrix. We assume that  $i(\theta)$  and third-order cumulants of $U(\theta)$ are finite and of order $O(n)$, where $n$ is the sample size or, more generally, an index  of information in the data about the model parameters.  We denote by $i_{rs}$ a generic entry of $i(\theta)$ and $i^{rs}$ an entry of its inverse, $i(\theta)^{-1}$ $(r,s=1,\ldots,p)$. Let 
$P_r(\theta)=E\{U(\theta)U(\theta)^\top U_r(\theta)\}$ and $Q_r(\theta)=-E\left\{j(\theta)  U_r(\theta) \right\}$ $(r=1,\ldots,p)$ be  $p\times p$ matrices of expected values  of log likelihood derivatives. Let, in addition, $[C]_r$ be the $r$th  column of matrix $C$.
 
The modified score equation proposed by \citet[][equation (10)]{kenne2017}  has the form
%\begin{equation}\label{modscoreeq2}
% \bar U_r +M_{r}=0 \quad (r=1,\ldots,p),
%%\,\,\text{ with } M_{r}=-\kappa_{1r}+\frac{1}{6} \frac{\kappa_{3r}}{\kappa_{2r} },\quad(r=1,\ldots,p), 
% \end{equation}
% where $\bar U_r $ is the efficient score, $\bar U_r=U_r-\sum_{a\neq r} \gamma^r_aU_a$,  with  $\gamma^r_a$ ($a=1,\ldots,p$, $a\neq r$)  the multiple regression coefficients of $U_r$ on the remaining components of $U(\theta)$, while $M_r$ is an adjustment term of order $O(1)$, involving  the  approximate first three cumulants of the profile score for $\theta_r$.   
%The expression of $M_r$ requires multiple summation, whose direct implementation is highly demanding in terms of computational time, especially for large $p$. 
 
%Here we show that (\ref{modscoreeq2}) is equivalent to 
\begin{align}\label{modscoreeq}
\tilde U(\theta)=U(\theta)+  \tilde A(\theta)=0\,,
\end{align}
with   $ \tilde  A(\theta)=i(\theta)M_1(\theta)$, where $M_1(\theta)$ is an adjustment term of order O(1) which requires multiple summation, whose direct implementation is computationally demanding, especially for large $p$.  

We show that  the $r$-th component of the  $p$-dimensional vector  $M_1(\theta)$  can be written as
\begin{align}\label{m1}
M_{1r} =[i(\theta)^{-1}]_r^\top\left( \frac{1}{2}F_1-F_{2,r}\right)\quad (r=1,\ldots,p),
\end{align}
where  $F_1$ and $F_{2,r}$ are $p$-dimensional vectors having entries
\begin{eqnarray*}
F_{1s} &=& \tr\left[i(\theta)^{-1}\{P_s(\theta)+Q_s(\theta)\}\right], \\
F_{2s,r} &=& \tr\left[h_r(\theta)\{(1/3)P_s(\theta)+(1/2)Q_s(\theta)\}\right]\quad (s=1,\cdots,p)\,,
\end{eqnarray*}
%$$
%F_{1l} = \tr\left[i(\theta)^{-1}\{P_l(\theta)+Q_l(\theta)\}\right]\text{ and }  F_{2l,r} = \tr\left[h_r(\theta)\{(1/3)P_l(\theta)+(1/2)Q_l(\theta)\}\right],
%$$
with  $h_r(\theta)=[i(\theta)^{-1}]_r[i(\theta)^{-1}]_r^\top/i^{rr}(\theta)$ a $p\times p$ matrix. 

The main steps for the derivation of (\ref{m1})  are as follows.  
Let   $\nu_{s,t,u}=E(U_sU_tU_u)$ and $\nu_{s,tu}=E(U_sU_{tu}) $ $(t,u=1,\cdots,p)$,
 %be the $(t,u)$ entries of $P_s$ and $Q_s$ in (\ref{modscoreeq}) for fixed $s$, respectively, 
 where  $U_{tu}=\partial U_t/\partial\theta_u$. 
In the following,  we adopt the Einstein summation convention, i.e., summation is implied over repeated indices $a,b,c,d$ taking values in $\{1,\ldots,p\}\setminus \{r\}$, with $r$ being any index in $\{1,\ldots,p\}$. The Einstein summation convention does not affect the index $r$. 
The quantity $M_1(\theta)$ in (\ref{modscoreeq}) has $r$-th component $M_r/\kappa_{2r}$ with
\begin{equation}\label{modscoreindex}
M_{r}=-\kappa_{1r}+\frac{1}{6} \frac{\kappa_{3r}}{\kappa_{2r} }\quad(r=1,\ldots,p), 
 \end{equation}
 where  $\kappa_{1r}, \kappa_{2r}$ and $\kappa_{3r}$ are  terms of order $O(1),O(n)$ and $O(n)$, respectively, and represent the  approximate first  three cumulants of the profile score for $\theta_r$. These cumulants are    given by
 \begin{align*}
 \kappa_{1r} &= -\frac{1}{2}\nu^{ab}\{(\nu_{r, ab}-\gamma_{d}^r\nu_{d,ab})+(\nu_{r, a,b}- \gamma_{d}^r\nu_{a,b,d})\},
 \quad \kappa_{2r}  = i_{rr}-\gamma_{a}^ri_{ra},\\
\kappa_{3r} & = \nu_{r,r,r}-3\gamma_{a}^r\nu_{r,r, a}+3\gamma_{a}^r \gamma_{b}^r\nu_{r, a,b }-\gamma_{a}^r\gamma_{b}^r \gamma_{c}^r\nu_{a,b,c},
 \end{align*}
 where $\gamma_{a}^r=i_{rb} \nu^{ab}$, with $\nu^{ab}$ a generic entry of the inverse of the matrix $i$ with entries $i_{ab}$.
  In  (\ref{modscoreindex}), all quantities are evaluated at $\theta$.  
  Using  block matrix inversion of $i(\theta)$, we have  $\gamma_{a}^r=-i^{ra}/i^{rr}$. After some algebra,  the  cumulants can be rewritten as 
 \begin{align*}
   \kappa_{1r} &= -\frac{1}{2}i^{rs}\nu_r^{tu}(\nu_{s,tu}+\nu_{s,t,u})/i^{rr},\quad \nu_r^{tu}=i^{tu}-i^{tr}i^{ru}/i^{rr}, \\
     \kappa_{2r} &= 1/i^{rr}, \quad \kappa_{3r} =i^{rs}i^{rt}i^{ru} \nu_{s,t,u}/(i^{rr})^3.
 \end{align*}
 In the above, the Einstein summation convention  applies to  indices $s,t,u$,   taking values in $\{1,...,p\}$. 
  Therefore, with some algebra, we can write $M_1(\theta)=K_1(\theta)/2+K_2(\theta)/6$, where  $K_1(\theta)$ and $K_2(\theta)$  are vectors with  generic entries
$$
 K_{1r}= i^{rs}\tr\left\{ \nu_{\theta,r}(P_s(\theta)+Q_s(\theta) \right\},\,\,
  K_{2r}= i^{rs}\tr\left\{ h_r(\theta) P_s(\theta) \right\}\,\, (r=1,\ldots,p),
$$
 where $\nu_{\theta,r}=i(\theta)^{-1}-h_r(\theta)$, with $h_r(\theta)=[i(\theta)^{-1}]_r[i(\theta)^{-1}]_r^\top/i^{rr}$. 
 We note that $P_s$ and $Q_s$ are $p\times p$ matrices with  $(t,u)$ entries  $\nu_{s,t,u}$ and $\nu_{s,tu} $, respectively.
It is then straightforward to see that    $M_1(\theta)$  has generic component (\ref{m1}).\\ 

\noindent
{\it Remark 1.}
The new expression of median bias adjustment based on (\ref{m1}) has already been applied in \citet{kks2019}  in the particular case of  normal meta regression models, but its general derivation is  given here. \\

\noindent
{\it Remark 2.}
 The adjustment term $\tilde A(\theta)$ can   be written as 
 $$
 \tilde A(\theta)=A^*(\theta)-i(\theta)\tilde F_2\,, 
$$
where $A^*(\theta)=(1/2)F_1$, with the vector $\tilde F_2$ having entries  
$\tilde F_{2r}=[i(\theta)^{-1}]_r^\top F_{2,r}$ $(r=1,\ldots,p)$.  The term $A^*(\theta)$ equals  the mean bias adjustment of \citet{firth1993} as given in \citet[][formula (2.2)]{kosmi2010}. The additional term $-i(\theta)\tilde F_2$
is required for median bias reduction.  This highlights a general connection between mean and median bias reduction, that was noted 
for generalized linear models in  \citet{kosmidis2018mean}. Indeed, in such models the general expression of the adjustment term $\tilde A(\theta)$ given above simplifies in the compact form provided by \citet[][formula (9)]{kosmidis2018mean}. Details are given in the Appendix.\\

\noindent
{\it Remark 3.}
Letting $\hat\theta$ be the maximum likelihood estimate,
   $\hat\theta^*$ the mean bias-reduced estimate of \citet{firth1993} and $\tilde\theta$ the median bias-reduced estimate  obtained as a solution of (\ref{modscoreeq}),
 a quasi-Fisher scoring-type algorithm has  $k$th iteration
$$\theta^{(k+1)}=\theta^{(k)} + i^{-1}(\theta^{(k)})B(\theta^{(k)})+i^{-1}(\theta^{(k)})U(\theta^{(k)}),$$
where $B(\theta)$ can be the null vector,  $A^*(\theta)$  or $\tilde A(\theta)$, giving as a solution $\hat\theta$, $\hat\theta^*$ or $\tilde\theta$, respectively.  In the light of Remark 2 above, the updating term for median bias reduction is the one for mean bias reduction with the additional term $- \tilde F_2(\theta^{(k)})$.

\section{Double index beta regression}
%%%%%%%%%%%%%%%%%%%%%%%%%%%%%%

%The beta regression is a model for continuous response variables  that are measured on the open %interval (0, 1). The  beta  distribution is flexible for modelling rates and proportions since its density has %quite different shapes. Among particular forms, its distribution could be unimodal, uniform, or bimodal. 
%%%%%%%%%%%%%%%%%%%%%%%%%%%%%%
%\subsection{Model}
%%%%%%%%%%%%%%%%%%%%%%%%%%%%%%
The double index beta regression model was introduced  by \citet{smith2006} and \citet{sima2010} as a generalization of beta regression \citep{ferra2004} in order to account for covariate effects both on the mean and on the precision parameter when modelling continuos rates and proportions. An {\tt R} implementation of mean bias reduction was developed in \citet{grun2012}, while median bias reduction for the case with constant precision was considered  in \citet[][Example 8]{kenne2017}. Simulation results in the latter paper show that both mean and median bias reduction have considerable success in achieving the respective goals,   for estimation of the precision parameter, as compared to maximum likelihood. On the other hand, all methods show similar  accurate behaviour for estimation of mean parameters, with mean and median bias reduction providing nevertheless improved coverage of confidence intervals.  
Here we extend median bias reduction to the double index case. 

Let $y_1,\ldots,y_n$ be  realizations of independent random variables $Y_1,\ldots,Y_n$, each having a beta distribution with parameters $\phi_i\mu_i$ and $\phi_i(1-\mu_i)$, i.e.\ with expected value $\mu_i$ and precision parameter $\phi_i$. The density of $Y_i$ is
\begin{align}\label{dens}
f_{Y_i}(y_i;\mu_i,\phi_i)=\frac{\Gamma(\phi_i)}{\Gamma(\mu_i\phi_i)\Gamma\{(1-\mu_i)\phi_i\}}y_i^{\mu_i\phi_i-1}(1-y_i)^{(1-\mu_i)\phi_i-1},\, 
\end{align}
where $0<y_i<1$, $0<\mu_i<1$, $\phi_i>0$, and $\Gamma(\cdot)$ is the gamma function. 
%From this parameterization, it follows that the random variable $Y$ has mean $E(Y)=\mu$ and  variance $Var(Y)=\mu(1-\mu)/(1+\phi)$. We note that the variance, for fixed $\mu$, decreases as $\phi$ increases and then, the parameter $\phi$ can be seen as a precision parameter. 
Double index beta regression  assumes a  regression structure  both for the expected value $\mu_i=g_1^{-1}(\eta_i)$ and for the precision $\phi_i=g_2^{-1}(\zeta_i)$, 
where $\eta_i= x_i \beta$  and $\zeta_i=z_i\gamma$, with  $x_i=(x_{i1},\ldots, x_{ip})$  and $z_i=(z_{i1},\ldots, z_{iq})$ representing row vectors of covariates. Above,  $\beta=(\beta_1,\ldots,\beta_p)^\top$ and $\gamma=(\gamma_1,\ldots,\gamma_q)^\top$ are  vectors of unknown regression parameters.
Additionally,  the link functions $g_1(\cdot)$ and $g_2(\cdot)$ are monotonic and should have the  mapping property $g_1:(0,1) \rightarrow \R$ and $g_2:(0,\infty) \rightarrow \R$, respectively. Obvious choices for $g_1(\cdot)$ include the  logit and  probit; and  for $g_2(\cdot)$ the log, the square root and the identity, with only the log satisfying the mapping property. 

%Unlike  with generalized linear models, a compact simplified expression for the adjustment is not available for the %current model. 
Here $\theta = (\beta^\top, \gamma ^\top)^\top$ and the ingredients $P_s(\theta)$ and $Q_s(\theta)$ required for  (\ref{m1}) are given in the Appendix. They are obtained along the lines of  \citet[][Section 2.3]{grun2012}, who gave the quantities $P_s(\theta) + Q_s(\theta)$  for mean bias reduction.   

%The use of (\ref{modscoreeq}) for inference  about $\theta=(\beta_1,\ldots,\beta_p,\gamma_1,\ldots,\gamma_q)^\top$ extends the example 8 in \citet{kenne2017} in which a  regression structure is assumed   only on the expected value. Here, we again consider  the latter model together with the current one focusing attention to the invariance property of the estimators.   

%%%%%%%%%%%%%%%%%%%%%%%%%%%%%%
%\subsection{Iterative median bias reduction}
%%%%%%%%%%%%%%%%%%%%%%%%%%%%%%

%%%%%%%%%%%%%%%%%%%%%%%%%%%%%%
%\subsection{Simulation experiments}
%%%%%%%%%%%%%%%%%%%%%%%%%%%%%%

We use Monte Carlo simulation to assess the performance of the median bias reduced estimator. For this purpose, we consider a model having a regression structure on both mean and precision, with logit and log link, respectively. In particular, we let
\begin{equation}\label{modelexp2}
\begin{split}
\log \frac{\mu_i}{1-\mu_i} &= \beta_0+\beta_1x_{i1} + \beta_2x_{i2},\\
\log \phi_i &= \gamma_0+\gamma_1x_{i1} + \gamma_2x_{i2}\quad (i = 1,\ldots,n),\\ 
\end{split}
\end{equation}  
where the $x_{i1}$  are $n$ independent realizations of a standard normal and $x_{i2}=\log u_i$, with $u_i$ generated from a  uniform distribution on $(1,2)$ ($i=1,\ldots,n$).
 The values of the parameters are $\beta_0=1.5,\, \beta_1=0.5,\, \beta_2=2$,  
$\gamma_0=1.7,\, \gamma_1=0.7$ and $ \gamma_2=3$, chosen so as to have average simulated  proportions around  0.9.

%We generated 10000 simulated samples for $n=20,\,40$ and 60.

 % with similar settings as in \citet{sima2010}.  The first one considers  a logit link on the mean structure  without  linear dispersion covariates as 
%\begin{equation}\label{modelexp1}
%\log \frac{\mu_i}{1-\mu_i}=\beta_0+\beta_1x_{i1} + \beta_2x_{i2},\quad i = 1,\ldots,n,
%\end{equation} 
%where the $x_{i1}$ are $n$ realizations of a standard normal and $x_{i2}=\log u_i$, with $u_i$ generated from a  uniform $U(1,2)$, $i=1,\ldots,n$. For the precision parameter, both the $\tau=\log\phi$ parameterization  and the $\phi$ parameterization have been considered.  
%Parameter values were fixed as  $\beta_0=1.5,\, \beta_1=0.5,\, \beta_2=2$ and $\phi=200$.  

The sample sizes considered were $n=20,\,40$ and 60. For each $n$,
we run    10000  Monte Carlo replications, where the values of the explanatory variables  $x_{i1}$ and $x_{i2}$    were held constant throughout the simulations. For each replication, the model was fitted using maximum likelihood, mean bias reduction and median bias reduction. We summarize the simulation results through the  estimated percentage probability of underestimation (PU), estimated bias (BIAS), root mean square error (RMSE) and percentage estimated coverage probability of 95\% Wald-type confidence intervals (WALD).

The maximum likelihood and mean  bias reduced estimates were obtained from the \texttt{R} package \texttt{betareg} \citep{grun2012}, while the median bias-reduced estimates were calculated from the \texttt{R} function \texttt{mbrbetareg} \citep{kpal2017b}.

\begin{table}[!htbp]
\caption{Simulation results for double index beta regression.  }\label{exp21}
\centering
\vspace{0.1cm}
\centering
\begin{tabular}{lllcccccc}
& & & $\beta_0$ & $\beta_1$ & $\beta_2$ & $\gamma_0$ & $\gamma_1$ & $\gamma_2$ \\ 
  \hline							
\multirow{12}{*}{$n=20$}  &PU 	&  $\hat\theta$ 	&45.9	&50.5	&52.0	&31.7	&46.9	&54.8\\
  & &	 $\hat\theta^*$ 	&54.7	&49.9	&47.2	&57.5	&49.8	&44.5\\
  & &	  $\tilde\theta$ 	&51.3	&50.4	&49.1	&52.6	&50.6	&47.5\\
   \cline{3-9} 
   &BIAS &	 $\hat\theta$ 	&0.07	&-0.01&	-0.07	& 0.54&	0.03	&-0.35\\
   & &	 $\hat\theta^*$ 	&-0.04	&0.00	&0.05	&-0.08	&-0.01	&0.16\\
  & &	 $\tilde\theta$ 	&0.01	&-0.01	&0.00	&0.04	&-0.02	&0.02\\
 \cline{3-9} 							
&  RMSE &	 $\hat\theta$ 	&0.51	&0.22	&0.94	&1.15	&0.52	&2.07\\
  & &	 $\hat\theta^*$ 	&0.50	&0.21	&0.93	&0.95	&0.45	&1.87\\
 	& & $\tilde\theta$ 	&0.50	&0.21	&0.92	&0.94	&0.45	&1.85\\
 \cline{3-9} 							
& WALD 	& $\hat\theta$ 	&82.9	&82.4	&83.3	&80.5	&79.8	&84.0\\
 & &	 $\hat\theta^*$ 	&90.3	&88.8	&90.7	&86.7	&85.0	&86.9\\
& & 	 $\tilde\theta$ 	&89.5	&88.5	&90.0	&87.5	&85.1	&87.3\\
\hline

\multirow{12}{*}{$n=40$}  &PU  &	 $\hat\theta$ 	&47.1	&50.8	&51.1	&38.9	&44.3	&49.5\\
 & &	 $\hat\theta^*$ 	&51.7	&50.2	&49.4	&52.8	&49.5	&48.8\\
 & &	 $\tilde\theta$ 	&50.3	&50.2	&49.9	&50.5	&49.7	&49.0\\
  \cline{3-9} 	
 &BIAS &	 $\hat\theta$ 	&0.03	&0.00	&-0.01	&0.17	&0.03	&0.01\\
 & &  	 $\hat\theta^*$ 	&0.00	&0.00	&0.01	&-0.01	&0.00	&0.03\\
 & &	 $\tilde\theta$ 	&0.01	&0.00	&0.01	&0.02	&0.00	&0.02\\
 \cline{3-9} 							
  &RMSE &	 $\hat\theta$ 	&0.29	&0.12	&0.64	&0.55	&0.25	&1.28\\
 & &	 $\hat\theta^*$ 	&0.29	&0.12	&0.63	&0.51	&0.23	&1.24\\
  & &	 $\tilde\theta$ 	&0.29	&0.12	&0.63	&0.51	&0.23	&1.24\\
 \cline{3-9} 							
& WALD 	&  $\hat\theta$ 	&91.0	&88.8	&90.0	&89.3	&86.9	&89.7\\
 & &	 $\hat\theta^*$ &	93.4	&91.6	&92.7	&91.3	&89.8	&91.0\\
 & &	 $\tilde\theta$ 	&93.2	&91.3	&92.4	&91.4	&89.8	&91.0\\
 \hline	
 							
 \multirow{12}{*}{$n=60$}  &PU 	& $\hat\theta$ 	&47.5	&50.6	&50.8	&41.0	&51.9	&50.2\\
 & &	 $\hat\theta^*$ 	&50.6	&48.4	&50.1	&51.2	&47.2	&50.3\\
 & &	 $\tilde\theta$ 	&49.4	&49.0	&50.6	&49.5	&48.5	&50.9\\	
  \cline{3-9} 					
 &BIAS &	 $\hat\theta$ 	&0.02	&-0.01	&-0.01	&0.11	&-0.02	&-0.01\\
  & &	 $\hat\theta^*$ 	&0.00	&0.00	&0.00	&0.01	&0.01	&-0.01\\
 & &	 $\tilde\theta$ 	&0.01	&0.00	&-0.01	&0.03	&0.00	&-0.02\\

 \cline{3-9} 							
 &RMSE &	 $\hat\theta$ 	&0.27	&0.12	&0.53	&0.45	&0.23	&0.96\\
  & &	 $\hat\theta^*$ 	&0.26	&0.12	&0.53	&0.43	&0.22	&0.94\\
 & &	 $\tilde\theta$ 	&0.27	&0.12	&0.53	&0.43	&0.22	&0.94\\
 \cline{3-9} 							
  &WALD &	 $\hat\theta$ 	&92.4	&91.4	&92.3	&91.7	&90.5	&92.1\\
 & &	 $\hat\theta^*$ 	&93.6	&93.0	&93.5	&93.0	&91.2	&92.6\\
 & &	 $\tilde\theta$ 	&93.4	&92.7	&93.3	&92.8	&91.3	&92.7\\

   \hline
\end{tabular}
\end{table}
The results are presented in Table \ref{exp21}. Both  median and  mean bias reduced estimators are superior to the maximum likelihood estimator. In particular, $\tilde\theta$ is effective in median centering, especially for $n=20,40$. In terms of mean bias reduction,  $\tilde\theta$  and  $\hat\theta^*$ are equivalent up to simulation error.  The two estimators have comparable coverages of confidence intervals. 

Covariate effects on the dispersion parameter are typically expressed in the reparameterization $\exp(\gamma_1)$ and $\exp(\gamma_2)$. Estimated bias of estimators $\exp(\hat\gamma_j)$, $\exp(\hat\gamma_j^*)$ and  $\exp(\tilde\gamma_j^*)$ ($j=1,2$)  are given in Table \ref{exp2bias}. Bias reduction is not expected to be effective for the transformed estimator $\exp(\hat\gamma_j^*)$  and, indeed, in most cases $\exp(\hat\gamma_j^*)$ does not show the smallest bias.

\begin{table}[ht]
\caption{Estimated bias of estimators $\exp(\hat\gamma_j)$, $\exp(\hat\gamma_j^*)$ and  $\exp(\tilde\gamma_j^*)$
in the model (\ref{modelexp2}).}\label{exp2bias}
\vspace{0.1cm}
\centering
\resizebox{1\textwidth}{!}{\begin{tabular}{l | ccc | ccc | ccc}
\multicolumn{1}{c}{} &\multicolumn{3}{c}{$n=20$} &\multicolumn{3}{c}{$n=40$} &\multicolumn{3}{c}{$n=60$} \\
   \cline{2-10} 
 & $\hat\theta$ & $\hat\theta^*$ &$\tilde\theta$ & $\hat\theta$ & $\hat\theta^*$ & $\tilde\theta$ & $\hat\theta$ & $\hat\theta^*$ & $\tilde\theta$ \\ 
  \hline
$\exp(\gamma_1)$ & 0.355 & 0.185 & 0.166 & 0.128 & 0.051 & 0.050 & 0.015 & 0.072 & 0.055 \\ 
  $\exp(\gamma_2)$ & 68.505 & 91.987 & 75.348 & 25.734 & 24.185 & 23.869 & 11.190 & 10.763 & 10.322 \\ 
   \hline
\end{tabular}}
\end{table}

\section{Beta-binomial regression}

The beta-binomial model is  popular  for analysing data in form of number of successes  in a given number of trials,  $y_i\in \{0,\ldots,m_i\}$. It is suitable in situations where extra-binomial variability  is present. The model is defined by assuming that 
 $Y_i$ conditionally on $\pi_i$ is binomial with index $m_i$ and success probability $\pi_i$  and that  $\pi_i$ has a beta distribution with shape parameters $\alpha_i$ and $\beta_i$,  with  $\alpha_i > 0,\,$ $\beta_i >0$ so that $E(\pi_i)=\mu_i=\alpha_i/(\alpha_i+\beta_i),\,$ and   $Var(\pi_i)=\mu_i(1-\mu_i)\phi_i$, where  $\phi_i= 1/(\alpha_i+\beta_i+1)$ ($i=1,\ldots,n$).  The random variable $Y_i$ is marginally distributed as beta-binomial  with mean  $E(Y_i)=m_i \mu_i$ and variance $Var(Y_i)=m_i\mu_i(1-\mu_i)\{1+\phi_i(m_i-1)\}$.
The probability mass function of $Y_i$ is 
\begin{equation}\label{bbmodel}
\begin{split}
%&f_{Y_i}(y_i;\mu_i,\phi_i) =  \binom{m_i}{y_i}\frac{\Gamma(\alpha_i+\beta_i)}{\Gamma(\alpha_i)\Gamma(\beta_i)}\frac{\Gamma(\alpha_i+y_i) \Gamma(\beta_i+m_i-y_i)}{\Gamma{(\alpha_i+\beta_i+m_i)}}\\
f_{Y_i}(y_i;\mu_i,\phi_i)=\binom{m_i}{y_i} \frac{\prod_{j=0}^{y_i-1}[(1-\phi_i)\mu_i+j\phi_i] \prod_{j=0}^{m_i-y_i-1} [(1-\mu_i)(1-\phi_i)+j\phi_i]}{\prod_{j=0}^{m_i-1}[(1-\phi_i)+j\phi_i]},
\end{split}
\end{equation}
where $y_i\in \{0,\ldots,m_i\}$. We consider a  regression structure both for the expected value and for the precision, i.e.\ $\mu_i=g_1^{-1}(\eta_i)$   and $\phi_i=g_2^{-1}(\zeta_i)$, 
where $\eta_i= x_i \beta$  and $\zeta_i=z_i\gamma$, with  $x_i=(x_{i1},\ldots, x_{ip})$  and $z_i=(z_{i1},\ldots, z_{iq})$ representing row vectors of covariates and $g_1(\cdot)$, $g_2(\cdot)$ suitable link functions.   The overall parameter is denoted by $\theta=(\beta^\top,\gamma^\top)^\top$, where  $\beta=(\beta_1,\ldots,\beta_p)^\top$ and $\gamma=(\gamma_1,\ldots,\gamma_q)^\top$ are  vectors of unknown regression parameters.

For  model (\ref{bbmodel}) we provide here mean and median bias reduction.
With constant mean and precision, a bias corrected estimator was developed by \citet{saha2005}. 
The quantities required for the  calculation of (\ref{m1}) are reported in the Appendix.  

As an example, we consider the  low-iron rat teratology dataset analysed in  \citet{liang1993} and 
available in the \texttt{R} package \texttt{VGAM}. The goal is  to study the effects of  dietary regimens on fetal development in laboratory rats. Fifty-eight female rats were put on iron-deficient diets and divided into four groups. Group 1 is the untreated (low-iron) group; group 2 received injections on day 7 or day 10 only; group 3 received injections on days 0 and 7 and group 4 received injections weekly. The rats were made pregnant, sacrificed 3 weeks later, and the total number of fetuses and the number of dead fetuses  in each litter were counted along with  the level
of mother's hemoglobin.
We assume model (\ref{bbmodel})  for the response $Y_i$ in the $i$th litter, with 
\begin{equation}\label{ratsmodel}
\log\frac{\mu_i}{1-\mu_i}=\beta_0+\beta_1 x_{i1}+ \beta_2 x_{i2}+\beta_3 x_{i3}+\beta_4 x_{i4}, \quad \phi_i =\phi\,\, (i=1,\dots,58).\\
\end{equation}
The covariates $x_{ij}$ ($j=1,2,3$) are indicator variables for the $(j+1)$th group, while  $x_{i4}$ is the level of mother's hemoglobin. 

Estimates $\hat\theta$, $\hat\theta^*$ and $\tilde\theta$ of parameters of model (\ref{ratsmodel})  are displayed in Table \ref{rats_estimates}, with (d1) corresponding to estimates computed on a  subset of the data with litter size less or equal to 11, while (d2) corresponds to estimates obtained with the  whole data set.
The maximum likelihood estimates are obtained using the function {\tt vglm} from the \texttt{R} package \texttt{VGAM}, while mean and  median bias-reduced estimates were calculated using the \texttt{R} package \texttt{brbetabinomial}, available on GitHub \citep{kpal2019}. 

\begin{table}[ht]
\centering
\caption{Low-iron rat teratology data. Estimates  (standard errors) of the parameters of model (\ref{bbmodel}). The label  (d1) indicates results  obtained from  the subset of the data with litter size less or equal to 11, while (d2) corresponds to  the  whole data set.}\label{rats_estimates}
\vspace{0.1cm}
%\resizebox{\textwidth}{!}

\resizebox{\textwidth}{!}{\begin{tabular}{|l|ccc|ccc|}
 \cline{2-7}
 \multicolumn{1}{l|}{}  & \multicolumn{3}{c|}{(d1)} & \multicolumn{3}{c|}{(d2)}\\ 
  
 \multicolumn{1}{l|}{}  & $\hat\theta$ & $\hat\theta^*$  & $\tilde\theta$ & $\hat\theta$ & $\hat\theta^*$  & $\tilde\theta$  \\ 
  \hline
$\beta_0$ & ~0.866  (1.130) & ~0.870  (1.128) & ~0.882  (1.141) & ~2.129  (0.847) & ~2.039  (0.853) & 2.055  (0.858) \\ 
  $\beta_1$ & -4.144  (1.441) & -3.793  (1.428) & -3.890  (1.449) & -2.440  (0.856) & -2.369  (0.867) & -2.394 (0.872) \\ 
  $\beta_2$ & -5.413  (2.070) & -4.803  (1.998) & -4.918  (2.028) & -2.837  (1.354) & -2.662  (1.343) & -2.716  (1.354) \\ 
  $\beta_3$ & -6.079  (2.978) & -5.402  (2.921) & -5.548 ( 2.963) & -2.287  (1.796) & -2.207  (1.809) & -2.244 (1.819) \\ 
  $\beta_4$ & ~0.172  (0.253) & ~0.151 (0.251) & ~0.157 (0.254) & -0.169  (0.173) & -0.157 (0.174) & -0.157 (0.175) \\ 
  $\phi$ & ~0.226 (0.087) & ~0.268 (0.090) & ~0.269 (0.092) & ~0.236 (0.059) & ~0.260 (0.060) & ~0.261  (0.061) \\ 
   \hline
\end{tabular}}
\end{table}

To assess the properties of the estimators, we performed a simulation study with sample
size and covariates as in the the two data sets (d1) and (d2) from the low-iron rat teratology data. For each of the two settings (d1) and (d2), we  considered 10000 replications with  parameter values equal to the maximum likelihood fit and covariates held fixed at the observed values. For each sample, we calculated the three estimates $\hat\theta$, $\hat\theta^*$ and $\tilde\theta$. Infinite estimates occurred using maximum likelihood with percentage frequencies 58\% for (d1) and  16\% for (d2), respectively, so that results for $\hat\theta$ should be judged accordingly. For detecting infinite estimates  the diagnostics in \citet{lesaffre1989} were adapted to the current model.

%\begin{table}[ht]
%\centering
%\begin{tabular}{rrrrrrrrrrrrr}
%  \hline
% & V1 & V2 & V3 & V4 & V5 & V6 & V7 & V8 & V9 & V10 & V11 & V12 \\ 
%  \hline
%mle & 2.129 & 0.847 & -2.440 & 0.856 & -2.837 & 1.354 & -2.287 & 1.796 & -0.169 & 0.173 & 0.236 & 0.059 \\ 
%  br & 2.039 & 0.853 & -2.369 & 0.867 & -2.662 & 1.343 & -2.207 & 1.809 & -0.157 & 0.174 & 0.260 & 0.060 \\ 
%  mbr & 2.055 & 0.858 & -2.394 & 0.872 & -2.716 & 1.354 & -2.244 & 1.819 & -0.157 & 0.175 & 0.261 & 0.061 \\ 
%  mle.1 & 0.866 & 1.130 & -4.144 & 1.441 & -5.413 & 2.070 & -6.079 & 2.978 & 0.172 & 0.253 & 0.226 & 0.087 \\ 
%  br.1 & 0.870 & 1.128 & -3.793 & 1.428 & -4.803 & 1.998 & -5.402 & 2.921 & 0.151 & 0.251 & 0.268 & 0.090 \\ 
%  mbr.1 & 0.882 & 1.141 & -3.890 & 1.449 & -4.918 & 2.028 & -5.548 & 2.963 & 0.157 & 0.254 & 0.269 & 0.092 \\ 
%   \hline
%\end{tabular}
%\end{table}

\begin{table}[!htbp]
\caption{Low-iron rat teratology data. Simulation of size $10^4$ of the  regression coefficient estimates with constant dispersion parameter  under the maximum likelihood fit. The label  (d1) indicates results  obtained from  the subset of the data with litter size less or equal to 11, while (d2) corresponds to  the  whole data. For maximum likelihood, the bias, root mean squared error and coverage are conditional upon finiteness of the estimates. }\label{rats}
\vspace{0.1cm}
\centering
\begin{tabular}{lllcccccc}
  \hline
& & & $\beta_0$ & $\beta_1$ & $\beta_2$ & $\beta_3$ & $\beta_4$& $\phi$ \\ 
  \hline

	\multirow{12}{*}{(d1)} &	 \multirow{3}{*}{PU} 	 &$\hat\theta$ 	&49.4	&55.8	&63.4	&63.5	&49.3	&66.8\\
	& & 	  $\hat\theta^*$  	&49.4	&46.2	&44.7	&45.3	&52.3	&49.1\\
 	& & 	  $\tilde\theta$ 	&49.2	&49.5	&50.1	&48.1	&51.0	&48.1\\
	 \cline{3-9} 	
 &	 \multirow{3}{*}{BIAS} &	 $\hat\theta$ 	&0.01	&-0.32	&0.13	&-0.12	&0.02	&-0.04\\
& &	  	 $\hat\theta^*$ 	&0.01	&0.01	&0.17	&0.12	&0.00	&0.01\\
 & &	 	 $\tilde\theta$ 	&0.01	&-0.13	&-0.08	&-0.10	&0.01	&0.01\\	
 \cline{3-9} 								
	& \multirow{3}{*}{RMSE}  &	 $\hat\theta$ 	&1.26	&1.67	&2.34	&3.34	&0.28	&0.10\\
 	& &  	 $\hat\theta^*$  	&1.16	&1.50	&1.99	&2.99	&0.26	&0.09\\
 	 & &	 $\tilde\theta$ 	&1.19	&1.58	&2.08	&3.09	&0.27	&0.09\\
 \cline{3-9} 								
	& \multirow{3}{*}{WALD}  &	$\hat\theta$ 	&92.4	&92.8	&93.3	&93.3	&92.6	&81.6\\
& &	 	 $\hat\theta^*$  &	94.8	&94.8	&95.4	&95.4	&95.0	&90.0\\
 & &	 	 $\tilde\theta$ 	&94.8	&94.9	&95.6	&95.4	&94.8	&91.2\\
 \hline															
  	\multirow{12}{*}{(d2)} &	\multirow{3}{*}{PU}  &	 $\hat\theta$ &	47.5	&52.5	&55.2	&53.4	&51.5	&64.3\\
  	& & 	 $\hat\theta^*$  &	50.7	&48.5	&47.2	&49.3	&49.9	&50.5\\
  	& & 	 $\tilde\theta$ 	&50.0	&49.7	&50.0	&50.6	&49.9	&49.7\\
	  \cline{3-9} 
	&	\multirow{3}{*}{BIAS} &	$\hat\theta$ &	0.06	&-0.09	&0.04	&-0.10	&-0.01	&-0.02\\
 & &	 	  $\hat\theta^*$  &	-0.01&	0.00	&0.03	&-0.01	&0.00	&0.00\\
  & &	 	 $\tilde\theta$ 	&0.01	&-0.03	&-0.12	&-0.06	&0.00	&0.00\\
  \cline{3-9} 								
  	 &\multirow{3}{*}{RMSE}  &	 $\hat\theta$ 	&0.89	&0.92	&1.31	&1.89	&0.18	&0.06\\
  	& & 	 $\hat\theta^*$  &	0.85	&0.87	&1.35	&1.83	&0.17	&0.06\\
  	& & 	 $\tilde\theta$ 	&0.86	&0.89	&1.48	&1.87	&0.18	&0.06\\
  \cline{3-9} 								
  	 &\multirow{3}{*}{WALD}  &	 $\hat\theta$ 	&93.8	&93.7	&95.2	&94.0	&93.7	&88.6\\
  	& & 	 $\hat\theta^*$  &	95.1	&94.8	&95.8	&95.1	&95.2	&92.3\\
  	& & 	 $\tilde\theta$ 	&95.1	&94.7	&96.0	&95.0	&95.1	&93.2\\

   \hline
\end{tabular}
\end{table}

The results are presented in Table \ref{rats}. The estimated bias, root mean squared error and coverage probability of confidence intervals based on $\hat\theta$  are conditional upon its finiteness. Although this favours $\hat\theta$, both $\hat\theta^*$ and $\tilde\theta$ are uniformly better. Especially for small samples (d1), the median bias reduced  estimator is superior in achieving median centering, and also in terms of mean bias for some coefficients.  
In the larger sample sizes setting (d2), both mean and median bias reduced estimators  achieve the desired goals and are uniformly preferable to maximum likelihood.

\appendix
\section*{Appendix}

\noindent
{\it $\tilde{A}(\theta)$ for generalized linear models}

%\begin{example} {\it Generalized linear models.}

\noindent
Let $y_1,\ldots,y_n$ be realizations  of independent  random variables $Y_1,\ldots,Y_n$, 
each with probability density or mass function of the exponential dispersion family form
\[
f_{Y_i}(y; \vartheta_i, \phi) = \exp\left\{\frac{y \vartheta_i - b(\vartheta_i) - c_1(y)}{\phi/m_i} - \frac{1}{2}a\left(-\frac{m_i}{\phi}\right) + c_2(y) \right\}
\]
for some sufficiently smooth functions $b(\cdot)$, $c_1(\cdot)$,
$a(\cdot)$ and $c_2(\cdot)$, and fixed observation weights
$m_1, \ldots, m_n$. The expected value of $Y_i$ is $E(Y_i) = \mu_i = b'(\vartheta_i)$  and its  variance is 
$Var(Y_i) = \phi b''(\vartheta_i)/m_i = \phi V(\mu_i)/m_i$.  Let $X$ be a $n\times p$ model matrix  
where each column corresponds to a predictor variable.
The model matrix $X$ is linked to $\mu_i$ through the relation $g(\mu_i)=\eta_i$ with $\eta_i=\sum_{r=1}^p\beta_r x_{ir}$ the linear predictor, where $x_{ir}$ is the $(i,r)th$ element of $X$, $\beta=(\beta_1,\ldots,\beta_p)^\top$ the regression coefficients and $g(\cdot)$ a monotone link function. For some models, the dispersion parameter, $\phi$ may also be estimated along with $\beta$.  We show below that the adjustment term
(\ref{m1}) gives, as a special case, the closed form expression obtained in \citet[Section 2.4]{kosmidis2018mean} for median bias reduction in generalized linear models.

 Let $\theta = (\beta^\top, \phi)^\top$ and $i_{\beta\beta}$ and $i_{\phi\phi}$ be the $(\beta,\beta)$ and  $(\phi,\phi)$ blocks of $i(\theta)$.  
Let $\diag\{\omega_i\}$ denote a diagonal matrix having $(\omega_1,\ldots,\omega_n)$ as its main diagonal. Let, in addition,  $0_p$ be a $p$-vector of zeros, $1_p$  a $p$-vector of ones and $0_{p\times p}$  a $p\times p$ matrix of zeros. 
The ingredients needed to calculate the adjustment term in (\ref{m1}) are as follows
\begin{align*}
& i(\beta,\phi) =
\left[
\begin{array}{cc}
i_{\beta\beta} & 0_p \\
0_p^\top & i_{\phi\phi}
\end{array}
\right]
=
\left[
\begin{array}{cc}
\frac{1}{\phi} X^\top W X & 0_p \\
0_p^\top & \frac{1}{2\phi^4}\sum_{i = 1}^n m_i^2 a''_i
\end{array}
\right]\,,\\
\phantom{o}\\
& P_s= \left\{ \begin{array}{l}
 \left[
\begin{array}{cc}
X^\top WO_{1s}X & X^\top WO_{2s}1_n \\
1_n^\top O_{2s}WX & 0
\end{array}
\right]\qquad\quad (s=1,\cdots,p),\\
\phantom{o}\\
   \left[
\begin{array}{cc}
 X^\top W X/\phi^2& 0_p\\ 
 0_p^\top & \frac{1}{2\phi^6}\sum_{i = 1}^n m_i^3 a'''_i\end{array}
\right]\quad (s=p+1),
\end{array} \right.\\
\end{align*}
\begin{align*}
&  Q_s=   \left\{ \begin{array}{l}
  \left[
\begin{array}{cc}
-X^\top W(O_{1s}-O_{3s})X & -X^\top WO_{2s}1_n \\
-1_n^\top O_{2s}WX & 0\end{array}
\right]\, (s=1,\ldots,p),\\
\phantom{o}\\
  \left[
\begin{array}{cc}
 0_{p\times p}& 0_p\\ 
 0_p^\top & - \frac{1}{\phi^5}\sum_{i = 1}^n m_i^2 a''_i\end{array}
\right], \qquad \qquad\quad\,\,\,\, (s=p+1),\\
\end{array} \right.\\
\phantom{o}\\
& F_{1s} =   \left\{
 \begin{array}{l} 2\sum_{i=1}^n x_{is}w_i\xi_i \qquad \,\, (s=1,\ldots,p), \\
 \phantom{o}\\
  \frac{p-2}{\phi}+\frac{ \sum_{i = 1}^n m_i^3 a'''_i}{\phi^2\sum_{i = 1}^n m_i^2 a''_i} \quad (s=p+1),\\
  \end{array} \right.\\
  \phantom{o}\\
& F_{2s,r} =\left\{
 \begin{array}{l} - \sum_{i=1}^n x_{is} \tilde{h}_{r,i} \left( \frac{d_iv'_i}{6v_i}-\frac{d'_i}{2d_i} \right)\quad  (r=1,\ldots,p,\,\,s=1,\ldots,p),\\
 \phantom{o}\\
 \frac{1}{3\phi} \sum_{i=1}^n \tilde{h}_{r,i} \qquad \qquad \qquad\quad\,\, (r=1,\ldots,p,\,\,s=p+1),\\
 \phantom{o}\\
  0  \qquad \qquad \qquad  \qquad \qquad \qquad (r=p+1,\,\,s=1,\ldots,p),\\
  \phantom{o}\\
 \frac{ \sum_{i = 1}^n m_i^3 a'''_i}{3\phi^2\sum_{i = 1}^n m_i^2 a''_i}-\frac{1}{\phi} \quad \qquad \qquad  (r=p+1,\,\,s=p+1),\\
\end{array} \right.\\
\end{align*}
where $W = {\rm diag}\left\{w_1, \ldots, w_n\right\}$, with
$w_i = m_i d_i^2/v_i$. Moreover, \ $O_{js} = {\rm diag}\left\{o_{js1},\ldots,o_{jsn}\right\}$ $(j=1,2,3)$, with $o_{1si}={x_{is}d_i v'_i}/(v_i \phi)$,
$o_{2si}=x_{is}/\phi^2$ and $o_{3si}={x_{is}d'_i}/(d_i \phi)$,
 $a''_i = a''(-m_i/\phi)$, $a'''_i = a'''(-m_i/\phi)$,
with  $a''(e) = d^2 a(e)/d e^2$ and  $a'''(e) = d^3 a(e)/d e^3$,  $d_i = d\mu_i/d\eta_i$,  $d_i' = d^2\mu_i/d\eta_i^2$,  $v_i = V(\mu_i)$,
$v'_i=V'(\mu_i)=d V(\mu_i)/ d\mu_i$,  $\xi_i = h_id_i'/(2d_iw_i)$, $h_i$ is
the `hat' value for the $i$th observation, obtained as the $i$th diagonal element of the matrix $H = X (X^\top W X)^{-1} X^\top W$, $\tilde{h}_{r,i}$ is the
$i$th diagonal element of $X G_r X^T W$, with
$ G_r = [i_{\beta\beta}^{-1}]_{r} [i_{\beta\beta}^{-1}]_{r}^\top / (\phi i^{rr})$.
Using all the above ingredients in (\ref{m1}), the  adjustments for $\beta$ and $\phi$  simplify to
\begin{equation}\label{adjglm}
 \tilde A_{\beta} =  X^\top W (\xi + X u) \quad \text{and} \quad \tilde A_\phi = \frac{p}{2\phi}+\frac{ \sum_{i = 1}^n m_i^3 a'''_i}{6\phi^2\sum_{i = 1}^n m_i^2 a''_i} \, ,
\end{equation}
respectively, where  $\xi =(\xi_1,\ldots,\xi_n )^\top$ and 
$u = (u_1, \ldots, u_p)^\top$ with
\begin{align*}
	u_r = [(X^\top W X)^{-1}]_{r}^\top X^\top \left[
	\begin{array}{c}
	\tilde{h}_{r,1} \left\{d_1 v'_1 / (6 v_1) - d'_1/(2 d_1)\right\} \\
	\vdots \\
	\tilde{h}_{r,n} \left\{d_n v'_n / (6 v_n) - d'_n/(2 d_n)\right\}
	\end{array}
	\right] \,.
\end{align*}
Expressions in (\ref{adjglm}) coincide with those  in \citet[Section 2.4]{kosmidis2018mean}. \\

% \end{example}

\noindent
{\it Quantities for median bias reduction in double index beta regression (Section 3).}

%We denote by  $\theta=(\beta^\top,\gamma^\top)^\top$ the overall vector parameters.  
%Typically, one estimates the parameters by maximizing 
For the beta regression model (\ref{dens}), the log likelihood function based on $n$ independent observations is 
\begin{align*}
\ell(\theta)=\sum_{i=1}^n \left[ \phi_i(1-\mu_i)s_i+ \phi_i\mu_it_i + \log\Gamma(\phi_i) - \log\Gamma(\phi_i\mu_i) - \log\Gamma\{\phi_i(1-\mu_i)\}  \right],
\end{align*}
where $s_i=\log(1-y_i)$ and $t_i=\log y_i$. Basic likelihood quantities needed for median bias reduction are the same as those derived by \citet{grun2012} for bias reduction. For ease of reference, they are given below. 
%From the above expression, we  note that $s_i$ and $t_i$ are sufficient statistics with natural parameters $ \phi_i(1-\mu_i)$ and $ \phi_i\mu_i$, respectively. 
The derivatives of the log likelihood with respect  to the $\beta$ and $\gamma$  are, respectively,
\begin{align}\label{score}
U_\beta =X^\top\Phi D_1(\tilde T-\tilde S)\, \text{ and   }\, U_\gamma = Z^\top D_2\{M(\tilde T-\tilde S)+\tilde S\}, 
\end{align}
where $\Phi=\diag\{\phi_1,\ldots,\phi_n\}$, $M=\diag\{\mu_1,\ldots,\mu_n\}$, $D_1=\diag\{d_{1,1},\ldots,d_{1,n}\}$ and $D_2=\diag\{d_{2,1},\ldots,d_{2,n}\}$, with 
$d_{1,i}=\partial\mu_i/\partial\eta_i$ and $d_{2,i}=\partial\phi_i/\partial\zeta_i$. The quantities $\tilde T$ and $\tilde S$ are the vectors of centered sufficient statistics, with $i$th component $\tilde T_i=t_i-E(T_i)$ and $\tilde S_i=s_i-E(S_i)$, respectively, where $E(T_i)=\psi^{(0)}(\phi_i\mu_i)-\psi^{(0)}(\phi_i)$ and $E(S_i)=\psi^{(0)}\{\phi_i(1-\mu_i)\}-\psi^{(0)}(\phi_i)$ with $\psi^{(l)}(a)=\partial^{l+1}\log\Gamma(a)/\partial a^{l+1}$  $(l=0,1,\ldots,i=1,\ldots,n)$. 
Moreover, $X$ and $Z$ denote the $n\times p$ and $n\times q$ design matrices with $i$th row $x_i$ and $z_i$ ($i=1,\ldots,n$), respectively. 
The Fisher information is given by 
$$
i(\theta)=
\left[
\begin{array}{cc}
i_{\beta\beta} & i_{\beta\gamma}\\
i_{\beta\gamma}^\top & i_{\gamma\gamma}
\end{array}  
 \right], 
$$
where 
\begin{align*}
i_{\beta\beta} &= X^\top D_1\Phi K_2\Phi D_1 X,\quad i_{\beta\gamma}=X^\top D_1\Phi(MK_2-\Psi_1)D_2Z,\\
 i_{\gamma\gamma}&= Z^\top D_2\{M^2K_2+(I_n-2M)\Psi_1-\Omega_1\}D_2Z.
\end{align*}
Above, $K_2=\diag\{\kappa_{2,1},\ldots,\kappa_{2,n}\}$, with $\kappa_{2,i}=Var(\tilde t_i-\tilde s_i)=\psi^{(1)}(\phi_i\mu_i)+\psi^{(1)}\{\phi_i(1-\mu_i)\}$ $(i=1,\ldots,n)$, 
$\Psi_l=\diag[\psi^{(l)}\{\phi_1(1-\mu_1)\},\ldots,\psi^{(l)}\{\phi_n(1-\mu_n)\} ]$, $\Omega_l=\diag\{\psi^{(l)}(\phi_1),\ldots,\psi^{(l)}(\phi_n) \}$   $(l=0,1,\ldots)$
 and $I_{n}$ is the $n\times n$ identity matrix. 
 
Using the observed information  available in \citet[formula (10)]{grun2012}, 
we obtain the quantities involved in (\ref{m1})
$$
P_s=
\left[
\begin{array}{cc}
V_{\beta\beta,s} & V_{\beta\gamma,s}\\
V_{\gamma\beta,s} & V_{\gamma\gamma,s}
\end{array}  
 \right] \text{ and }   
 Q_s=
\left[
\begin{array}{cc}
V^{'}_{\beta\beta,s} & V^{'}_{\beta\gamma,s}\\
V^{'}_{\gamma\beta,s} & V^{'}_{\gamma\gamma,s}
\end{array}  
 \right] \quad
 (s=1,\ldots,p), 
$$
where 
\begin{align*}
V_{\beta\beta,s} &= X^\top\Phi^3D^3_1K_3[X]_sX, \\
V_{\beta\gamma,s} &=V_{\gamma\beta,s}^\top = X^\top\Phi^2D_1^2D_2(MK_3+\Psi_2)[X]_sZ,\\
V_{\gamma\gamma,s} &= Z^\top\Phi D_1 D_2^2 (M^2K_3+2M\Psi_2-\Psi_2)[X]_sZ ,\\
V^{'}_{\beta\beta,s} &= X^\top\Phi^2D_1D^{'}_1K_2[X]_sX, \\
V^{'}_{\beta\gamma,s} &={V^{'}}_{\gamma\beta,s}^\top = X^\top\Phi D_1^2D_2K_2[X]_sZ,\\
V^{'}_{\gamma\gamma,s} &= Z^\top\Phi D_1D^{'}_2(MK_2-\Psi_1)[X]_s Z\,, 
\end{align*}
and 
$$
P_{p+t}=
\left[
\begin{array}{cc}
W_{\beta\beta,s} & W_{\beta\gamma,s}\\
W_{\gamma\beta,s} & W_{\gamma\gamma,s}
\end{array}  
 \right], \quad  
 Q_{p+t}=
\left[
\begin{array}{cc}
W^{'}_{\beta\beta,s} & W^{'}_{\beta\gamma,s}\\
W^{'}_{\gamma\beta,s} & W^{'}_{\gamma\gamma,s}
\end{array}  
 \right]\quad
 (t=1,\ldots,q), 
$$
where 
\begin{align*}
W_{\beta\beta,s} &= X^\top\Phi^2D_1^2D_2(MK_3+\Psi_2)[Z]_tX, \\
W_{\beta\gamma,s} &=W_{\gamma\beta,s}^\top = X^\top\Phi D_1D_2^2(M^2K_3+2M\Psi_2-\Psi_2)[Z]_tZ,\\
W_{\gamma\gamma,s} &= Z^\top D_2^3\{M^3K_3+(3M^2-3M+1_n)\Psi_2-\Omega_2\}[Z]_tZ ,\\
W^{'}_{\beta\beta,s} &= X^\top\Phi D_2D^{'}_1(MK_2-\Psi_1)[Z]_tX, \\
W^{'}_{\beta\gamma,s} &={W^{'}}_{\gamma\beta,s}^\top = X^\top D_1 D_2^2(MK_2-\Psi_1)[Z]_tZ,\\
W^{'}_{\gamma\gamma,s} &= Z^\top D_2D_2^{'}(M^2K_2+\Psi_1-2M\Psi_1-\Omega_1)[Z]_tZ,
\end{align*}
$D^{'}_1=\diag\{d^{'}_{1,1},\ldots,d^{'}_{1,n}\}$,\,$D^{'}_2=\diag\{d^{'}_{2,1},\ldots,d^{'}_{2,n}\}$, with 
$d^{'}_{1,i}=\partial^2\mu_i/\partial\eta_i^2$ and $d^{'}_{2,i}=\partial^2\phi_i/\partial\zeta_i^2$,
$K_3=\diag\{\kappa_{3,1},\ldots,\kappa_{3,n}\}$, with $\kappa_{3,i}=E\left\{(\tilde t_i-\tilde s_i)^3\right\}=\psi^{(2)}(\phi_i\mu_i)-\psi^{(2)}\{\phi_i(1-\mu_i)\}$ $(i=1,\ldots,n)$.\\

\noindent
{\it Quantities for mean and median bias reduction in beta-binomial regression (Section 4).}

For the beta-binomial model (\ref{bbmodel}), the  log likelihood function is
$$\ell(\theta)=\sum_{i=1}^n \ell(\theta;y_i)$$
with
$$\ell(\theta;y_i)= \sum_{j=0}^{y_i-1} \log\{E_{ij}\} +\sum_{j=0}^{m_i-y_i-1}\log\{F_{ij}\} - \sum_{j=0}^{m_i-1}\log\{G_{ij}\},$$
where 
\begin{align*}
E_{ij} &= (1-\phi_i)\mu_i+j\phi_i\\
F_{ij} & = (1-\mu_i)(1-\phi_i)+j\phi_i\\
G_{ij} &= (1-\phi_i)+j\phi_i.
\end{align*}
With constant dispersion, $\phi_i=\phi$ ($i=1,\ldots,n$), the log likelihood is the same as in 
\citet[][Appendix A]{saha2005}, where, however, $G_{ij}=G_j$ should be equal to $(1-\phi)+j\phi$. 

We have
%$U_{\mu_i} = (\partial/\partial \mu_i)\ell(\theta;y_i)$, 
%$U_{\phi_i} = U_{\phi_i}(\theta;y_i)=(\partial/\partial \phi_i(\ell(\theta;y_i)$, 
%$U_{\mu_i\mu_i} = U_{\mu_i\mu_i} (\theta;y_i)=(\partial^2/\partial \mu_i^2)\ell(\theta;y_i)$,
%$U_{\mu_i\phi_i} = U_{\mu_i\phi_i} (\theta;y_i)=(\partial^2/(\partial \mu_i \partial\phi_i))\ell(\theta;y_i)$,
%$U_{\phi_i\phi_i} = U_{\phi_i\phi_i} (\theta;y_i)=(\partial^2/\partial \phi_i^2)\ell(\theta;y_i)$.
%We have
 \begin{align*}
& U_{\mu_i}  = \frac{\partial \ell(\theta;y_i)}{\partial \mu_i}=\sum_{j=0}^{y_i-1} \frac{(1-\phi_i)}{E_{ij}} - \sum_{j=0}^{m_i-y_i-1} \frac{(1-\phi_i)}{F_{ij} },\\
 &U_{\phi_i} =\frac{\partial\ell(\theta;y_i)}{\partial \phi_i}=\sum_{j=0}^{y_i-1} \frac{(j-\mu_i)}{E_{ij}} + \sum_{j=0}^{m_i-y_i-1} \frac{(j+\mu_i-1)}{F_{ij} }-\sum_{j=0}^{m_i-1}\frac{(j-1)}{G_{ij}},\\
  & U_{\mu_i\mu_i} =  \frac{\partial^2\ell(\theta;y_i)}{\partial \mu_i^2}= -(1-\phi_i)^2\left[\sum_{j=0}^{y_i-1} \frac{1}{E_{ij}^2} + \sum_{j=0}^{m_i-y_i-1} \frac{1}{F_{ij}^2 }\right],\\
   & U_{\mu_i\phi_i} = \frac{\partial^2\ell(\theta;y_i)}{\partial \mu_i \partial\phi_i}=- \sum_{j=0}^{y_i-1} \frac{j}{E_{ij}^2} + \sum_{j=0}^{m_i-y_i-1} \frac{j}{F_{ij}^2 },\\
%\end{align*}
%\begin{align*}
    &U_{\phi_i\phi_i} = \frac{\partial^2\ell(\theta;y_i)}{\partial \phi_i^2}=-\sum_{j=0}^{y_i-1} \frac{(\mu_i-j)^2}{E_{ij}^2} - \sum_{j=0}^{m_i-y_i-1} \frac{(\mu_i+j-1)^2}{F_{ij}^2 }+\sum_{j=0}^{m_i-1}\frac{(j-1)^2}{G_{ij}^2}.\\
\end{align*}
 The derivatives of the log likelihood
function with respect to $\beta$ and $\gamma$ are
 $U_\beta=X^\top D_1^\mu U_\mu$ and  $U_\gamma=Z^\top D_1^\phi U_\phi$,  respectively, with $U_\mu=(U_{\mu_1},\ldots,U_{\mu_n})^\top,\,$ $U_\phi=(U_{\phi_1},\ldots,U_{\phi_n})^\top$ and $D_1^\mu=\diag\{d_{1,1}^\mu,\ldots,d_{1,n}^\mu\},\,$
 $D_1^\phi=\diag\{d_{1,i}^\phi,\ldots,d_{1,n}^\phi\},\,$ where $d_{1,i}^\mu=\partial\mu_i/\partial\eta_{1i}$ and $d_{1,i}^\phi=\partial\mu_i/\partial\eta_{2i}$.
 The second order partial derivatives are 
 \begin{align*}
 \frac{\partial^2 \ell(\theta)}{\partial\beta_r\partial\beta_s}&= \sum_{i=1}^n x_{ir}[U_{\mu_i\mu_i}(d_{1,i}^\mu)^2+U_{\mu_i}d_{2,i}^\mu]x_{is},\\
  \frac{\partial^2 \ell(\theta)}{\partial\beta_r\partial\gamma_a}&= \sum_{i=1}^n x_{ir}[U_{\mu_i\phi_i}d_{1,i}^\mu d_{1,i}^\phi]z_{ia},\\
   \frac{\partial^2 \ell(\theta)}{\partial\gamma_a\partial\gamma_b}&= \sum_{i=1}^n z_{ia}[U_{\phi_i\phi_i}(d_{1,i}^\phi)^2+U_{\phi_i}d_{2,i}^\phi]z_{ib},
 \end{align*}
 where $d_{2,i}^\mu=\partial^2\mu_i/\partial\eta_{1i}^2$ and $d_{2,i}^\phi=\partial^2\mu_i/\partial\eta_{2i}^2$. The expected Fisher information is given by
 $$
 i(\beta,\gamma) =
\left[
\begin{array}{cc}
i_{\beta\beta} & i_{\beta\gamma} \\
i_{\beta\gamma}^\top & i_{\phi\phi}
\end{array}
\right]
=
\left[
\begin{array}{cc}
-X^\top L_1(D_1^\mu)^2 X & -X^\top L_2 D_1^\mu D_1^\phi Z \\
-Z^\top  D_1^\phi D_1^\mu L_2 X  & -Z^\top L_3(D_1^\phi)^2 Z
\end{array}
\right],
 $$
where, for a generic subscript $j$, we let $L_j=\diag\{L_{j1},\ldots,L_{jn}\},\,$ and, in the formulae above,  
$$ 
L_{1i}=E(U_{\mu_i\mu_i})\,,\quad L_{2i}=E(U_{\mu_i\phi_i})\,,\quad 
L_{3i}=E(U_{\phi_i\phi_i})\,,
$$
with $E(U_{\mu_i\mu_i}) = \sum_{y=0}^{m_i} U_{\mu_i\mu_i}(\theta,y) P(Y_i=y)$.
Moreover, we let
\begin{align*}
&L_{4i}=E(U^3_{\mu_i})\,,\quad 
L_{5i}=E(U_{\mu_i}^2U_{\phi_i})\,,\quad 
L_{6i}=E(U_{\mu_i}U_{\phi_i}^2)\,,\quad 
L_{7i}=E(U_{\phi_i}^3)\\
&L_{8i}=E(U_{\mu_i\mu_i}U_{\mu_i})\,,\quad L_{9i}=E(U_{\mu_i}^2)\,,\quad 
L_{10i}=E(U_{\mu_i}U_{\mu_i\phi_i}) \,,\quad L_{11i}=E(U_{\mu_i}U_{\phi_i\phi_i})\,,\\
&L_{12i}=E(U_{\mu_i}U_{\phi_i}) \,,\quad L_{13i}=E(U_{\phi_i}U_{\mu_i\mu_i}) \,,\quad
L_{14i}=E(U_{\phi_i}U_{\mu_i\phi_i}) \,,\\&L_{15i}=E(U_{\phi_i}U_{\phi_i\phi_i})\,,\quad 
L_{16i}=E(U_{\phi_i}^2) \,.
\end{align*}

%\begin{align*}
%L_{8i}&=E\{U_{\mu_i\mu_i}U_{\mu_i}\}= \sum_{y=0}^{m_i} U_{\mu_i\mu_i}(\theta,y)U_{\mu_i}(\theta,y) P(Y_i=y),\\
%L_{9i}&=E\{U_{\mu_i}^2\} = \sum_{y=0}^{m_i} U_{\mu_i}(\theta,y)^2P(Y_i=y),\\
%L_{10i}&=E\{U_{\mu_i}U_{\mu_i\phi_i}\} = \sum_{y=0}^{m_i}  U_{\mu_i}(\theta,y)U_{\mu_i\phi_i}(\theta,y) P(Y_i=y),\\
%L_{11i}&=E\{U_{\mu_i}U_{\phi_i\phi_i}\} = \sum_{y=0}^{m_i}  U_{\mu_i}(\theta,y)U_{\phi_i\phi_i}(\theta,y) P(Y_i=y),\\
%L_{12i}&=E\{U_{\mu_i}U_{\phi_i}\} = \sum_{y=0}^{m_i}  U_{\mu_i}(\theta,y)U_{\phi_i}(\theta,y) P(Y_i=y),\\
%\end{align*}
%
%where
%\begin{align*}
%L_{13i}&=E\{U_{\phi_i}U_{\mu_i\mu_i}\} = \sum_{y=0}^{m_i}  U_{\phi_i}(\theta,y)U_{\mu_i\mu_i}(\theta,y) P(Y_i=y),\\
%L_{14i}&=E\{U_{\phi_i}U_{\mu_i\phi_i}\} = \sum_{y=0}^{m_i}  U_{\phi_i}(\theta,y)U_{\mu_i\phi_i}(\theta,y) P(Y_i=y),\\
%L_{15i}&=E\{U_{\phi_i}U_{\phi_i\phi_i}\} = \sum_{y=0}^{m_i}  U_{\phi_i}(\theta,y)U_{\phi_i\phi_i}(\theta,y) P(Y_i=y),\\
%L_{16i}&=E\{U_{\phi_i}^2\} = \sum_{y=0}^{m_i} U_{\phi_i}(\theta,y)^2P(Y_i=y).\\
%\end{align*}  

The further ingredients for the calculation of (\ref{m1}) are
$$
P_{s}=
\left[
\begin{array}{cc}
V_{\beta\beta,s} & V_{\beta\gamma,s}\\
V_{\gamma\beta,s} & V_{\gamma\gamma,s}
\end{array}  
 \right], \quad  
 Q_{s}=
\left[
\begin{array}{cc}
V'_{\beta\beta,s} & V'_{\beta\gamma,s}\\
V'_{\gamma\beta,s}  & V'_{\gamma\gamma,s}
\end{array}  
 \right]
  (s=1,\ldots,p)\,, 
$$
with
\begin{align*}
V_{\beta\beta,s}&=  X^\top L_4 (D_1^\mu)^3  X^D_s X,\,\, V_{\beta\gamma,s}=V_{\gamma\beta,s}^\top= X^\top L_5 (D_1^\mu)^2 D_1^\phi  X^D_s Z,\\
 V_{\gamma\gamma,s}&=  Z^\top L_6D_1^\mu (D_1^\phi)^2  X^D_s Z,\,\, V'_{\beta\beta,s}=  X^\top [L_8 (D_1^\mu)^3+L_9 D_1^\mu D_2^\mu] X^D_s X,\,\, \\V'_{\beta\gamma,s}&= (V'_{\gamma\beta,s})^\top= X^\top L_{10} (D_1^\mu)^2 D_1^\phi  X^D_s Z,\\
 V'_{\gamma\gamma,s}&=  Z^\top [L_{11}D_1^\mu (D_1^\phi)^2 +L_{12}D_1^\mu D_2^\phi ]X^D_s Z,
\end{align*} 
where $C^D_s$  denotes the diagonal matrix having the $s$-th column of a matrix C as its main diagonal and $D_2^\mu=\diag\{d_{2,1}^\mu,\ldots,d_{2,n}^\mu\}$, 
 $D_2^\phi=\diag\{d_{2,i}^\phi,\ldots,d_{2,n}^\phi\}$. 
 
 Moreover,
$$
P_{p+t}=
\left[
\begin{array}{cc}
W_{\beta\beta,t} & W_{\beta\gamma,}\\
W_{\gamma\beta,t} & W_{\gamma\gamma,t}
\end{array}  
 \right], \quad  
 Q_{p+t}=
\left[
\begin{array}{cc}
W'_{\beta\beta,t} & W'_{\beta\gamma,t}\\
W'_{\gamma\beta,t} & W'_{\gamma\gamma,t}
\end{array}  
 \right]  (t=1,\ldots,q) \,,
$$
with
\begin{align*}
W_{\beta\beta,t}&= X^\top L_5 (D_1^\mu)^2 D_1^\phi  Z^D_t X,
  W_{\beta\gamma,t}=W_{\gamma\beta,t}^\top=  X^\top L_6D_1^\mu (D_1^\phi)^2  Z^D_t Z ,\\ W_{\gamma\gamma,t}&=  Z^\top L_7 (D_1^\phi)^3  Z^D_t Z,\,\, W'_{\beta\beta,t}= X^\top[ L_{13} (D_1^\mu)^2 D_1^\phi+L_{12} D_1^\phi D_2^\mu ]Z^D_t X,\\
  W'_{\beta\gamma,t}&=(W'_{\gamma\beta,t})^\top=  X^\top L_{14}D_1^\mu (D_1^\phi)^2  Z^D_t Z ,\,\, W'_{\gamma\gamma,t}=  Z^\top [L_{15} (D_1^\phi)^3+L_{16}D_1^\phi D_2^\phi  ]Z^D_t Z,
\end{align*}

Further needed simplifications are 
$$
P_{s}+Q_{s}=
\left[
\begin{array}{cc}
V_{\beta\beta,s} +V'_{\beta\beta,s} & V_{\beta\gamma,s}+V'_{\beta\gamma,s}\\
V_{\gamma\beta,s} +V'_{\gamma\beta,s} &V_{\gamma\gamma,s}+ V'_{\gamma\gamma,s}
\end{array}  
 \right]\, (s=1,\ldots,p), 
$$
$$
P_{s}/3+Q_{s}/2=
\left[
\begin{array}{cc}
V_{\beta\beta,s}/3 +V'_{\beta\beta,s}/2 & V_{\beta\gamma,s}/3+V'_{\beta\gamma,s}/2\\
V_{\gamma\beta,s}/3 +V'_{\gamma\beta,s}/2 &V_{\gamma\gamma,s}/3+ V'_{\gamma\gamma,s}/2
\end{array}  
 \right]\, (s=1,\ldots,p), 
$$
$$
P_{p+t}+Q_{p+t}=
\left[
\begin{array}{cc}
W_{\beta\beta,t} +W'_{\beta\beta,t} & W_{\beta\gamma,t}+W'_{\beta\gamma,t}\\
W_{\gamma\beta,t} +W'_{\gamma\beta,t} &W_{\gamma\gamma,t}+ W'_{\gamma\gamma,t}
\end{array}  
 \right]\, (t=1,\ldots,q), 
$$
$$
P_{p+t}/3+Q_{p+t}/2=
\left[
\begin{array}{cc}
W_{\beta\beta,t}/3 +W'_{\beta\beta,t}/2 & W_{\beta\gamma,t}/3+W'_{\beta\gamma,t}/2\\
W_{\gamma\beta,t}/3 +W'_{\gamma\beta,t}/2 &W_{\gamma\gamma,t}/3+ W'_{\gamma\gamma,t}/2
\end{array}  
 \right]\, (t=1,\ldots,q), 
$$
where
\begin{align*}
& V_{\beta\beta,s}+V'_{\beta\beta,s}= X^\top [(L_4  +L_8 )(D_1^\mu)^3+L_9 D_1^\mu D_2^\mu]X^D_s X,\\
& V_{\beta\beta,s}/3+V'_{\beta\beta,s}/2= X^\top [(L_4/3  +L_8 /3)(D_1^\mu)^3+L_9 D_1^\mu D_2^\mu/2]X^D_s X,\\
& V_{\beta\gamma,s}+V'_{\beta\gamma,s}= X^\top [(L_5  +L_{10} )(D_1^\mu)^2D_1^\phi]X^D_s Z,\\
& V_{\beta\gamma,s}/3+V'_{\beta\gamma,s}/2= X^\top [(L_5/3  +L_{10}/2 )(D_1^\mu)^2D_1^\phi]X^D_s Z,\\
& V_{\gamma\gamma,s}+V'_{\gamma\gamma,s}= Z^\top [(L_6  +L_{11} ) D_1^\mu(D_1^\phi)^2+L_{12} D_1^\mu D_2^\phi]X^D_s Z,\\
& V_{\gamma\gamma,s}+V'_{\gamma\gamma,s}= Z^\top [(L_6/3  +L_{11}/2 ) D_1^\mu(D_1^\phi)^2+L_{12} D_1^\mu D_2^\phi/2]X^D_s Z,\\
& W_{\beta\beta,t}+W'_{\beta\beta,t}= X^\top [(L_5  +L_{13} )(D_1^\mu)^2D_1^\phi+L_{12} D_1^\phi D_2^\mu]Z^D_t X,\\
& W_{\beta\beta,t}/3+W'_{\beta\beta,t}/2= X^\top [(L_5/3  +L_{13}/2 )(D_1^\mu)^2D_1^\phi+L_{12} D_1^\phi D_2^\mu/2]Z^D_t X,\\
& W_{\beta\gamma,t}+W'_{\beta\gamma,t}= X^\top [(L_6  +L_{14} )D_1^\mu (D_1^\phi)^2]Z^D_t Z,\\
& W_{\beta\gamma,t}/3+W'_{\beta\gamma,t}/2= X^\top [(L_6 /3 +L_{14} /2)D_1^\mu (D_1^\phi)^2]Z^D_t Z,\\
& W_{\gamma\gamma,t}+W'_{\gamma\gamma,t}= Z^\top [(L_7  +L_{15} ) (D_1^\phi)^3+L_{16} D_1^\phi D_2^\phi]Z^D_t Z,\\
& W_{\gamma\gamma,t}/3+W'_{\gamma\gamma,t}/2= Z^\top [(L_7/3  +L_{15}/2 ) (D_1^\phi)^3+L_{16} D_1^\phi D_2^\phi/2]Z^D_t Z\,,\\
 &(s=1,\ldots,p, t=1,\ldots,q).
\end{align*}

With constant dispersion, $\phi_i=\phi$ ($i=1,\ldots,n$), the above quantities give the expected likelihood quantities in \citet[][Appendix A, p.\ 3511]{saha2005}, where, however,   $C_{2i}$  should be  equal to    $-B_{2i}^{(1,2,0,1)}-B_{1i}^{(1,2,0,1)}+B_{4i}^{(1,2,0,1)}+B_{4i}^{(2,1,1,0)}$ and $C_{7i}$ should be  equal to  $-B_{2i}^{(1,1,0,0)}-B_{1i}^{(1,1,0,0)}+2B_{4i}^{(1,1,0,0)}$, with the correction $G_j=(1-\phi)+j\phi$ in all terms.

 \section{Acknowledgements}
Nicola Sartori and Alessandra Salvan  were supported by the University of Padova under grant BIRD185955.

\bibliographystyle{chicago}
\bibliography{kpss.bib}
\end{document}